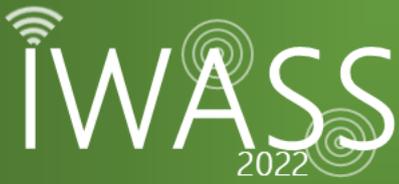

# IWASS
2022

International Workshop
on Autonomous
Systems Safety

# Proceedings

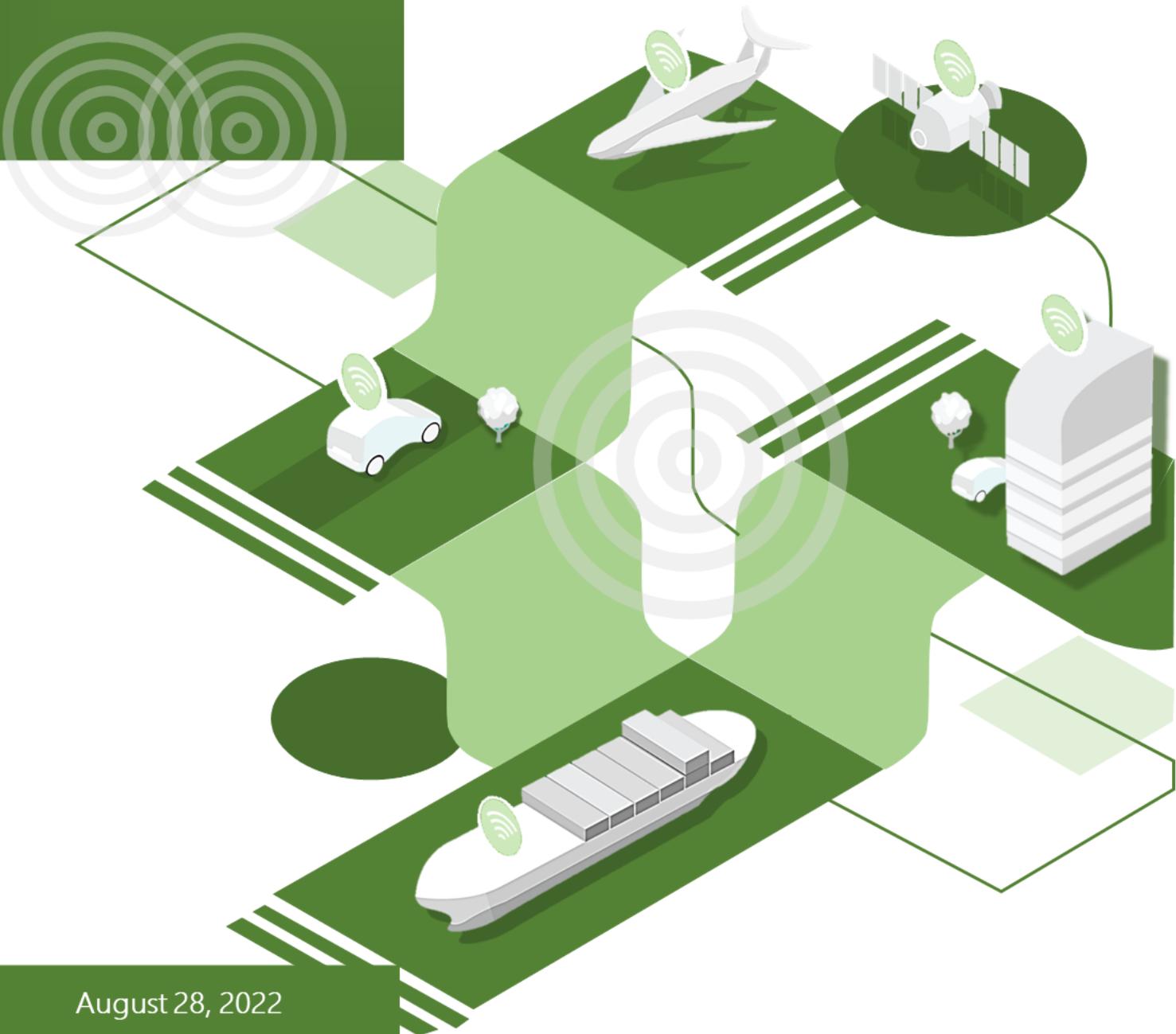

August 28, 2022

Dublin | Ireland

# Proceedings of the 3rd International Workshop on Autonomous Systems Safety

**Edited by:**

Christoph Thieme, Marilia Ramos, Ingrid B. Utne, Ali Mosleh

March 2023

UCLA ENGINEERING
B. John Garrick Institute for the Risk Sciences

NTNU – Trondheim
Norwegian University of
Science and Technology

DUBLIN
OLLSCOIL TEICNEOLAÍOCHTA
BHAILE ÁTHA CLIATH
TECHNOLOGICAL
UNIVERSITY DUBLIN

# Preface

The International Workshop for Autonomous System Safety (IWASS) is a joint effort by the B. John Garrick Institute for the Risk Sciences at the University of California Los Angeles (UCLA-GIRS) and the Norwegian University of Science and Technology (NTNU).

IWASS is an invitation-only event designed to be a platform for cross-industrial and interdisciplinary effort and knowledge exchange on autonomous systems' Safety, Reliability, and Security (SRS). The workshop gathers experts from academia, regulatory agencies, and industry to discuss challenges and potential solutions for SRS of autonomous systems from different perspectives. It complements existing events organized around specific types of autonomous systems (e.g., cars, ships, aviation) or particular safety or security-related aspects of such systems (e.g., cyber risk, software reliability, etc.). IWASS distinguishes itself from these events by addressing these topics together and proposing solutions for SRS challenges common to different types of autonomous systems.

IWASS 2022 was held on August 28th in Dublin, Ireland, and gathered 30 participants from 20 organizations from around the globe. In addition, a panel session at the European Safety and Reliability Conference (ESREL 2023) discussed the workshop's main conclusions and additional points with a larger audience.

This report summarizes IWASS 2022 discussions. It provides an overview of the main points raised by a community of experts on the current status of autonomous systems SRS. It also outlines research directions for safer, more reliable and secure future autonomous systems.

THIS PAGE INTENTIONALLY LEFT BLANK

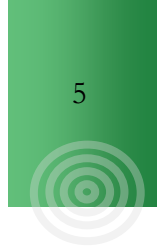

# Table of Contents

# Introduction

IWASS 2022 is the third edition of the workshop series on Autonomous System Safety, Reliability, and Security, initiated in 2019. The first IWASS was organized in 2019 in Trondheim, Norway. The 2019 event counted nearly 50 participants from different backgrounds, representing a diversity of industries, and from eight different countries. The proceedings published by NTNU[1] summarizes the discussions held at the workshop, briefly described in the next section, in addition to six research papers on autonomous systems SRS.

Initially planned as an in-person event in Los Angeles in 2020, IWASS 2021 switched to an online event in 2021 due to the COVID-19 pandemic and related travel restrictions. IWASS 2021 assembled a broad and diverse field of experts with 49 participants from 39 different organizations and nine countries. The workshop program was distributed over three days and included domain experts' presentations and discussion sessions, summarized in the proceedings published by UCLA-GIRS[2].

The third IWASS was organized as an in-person one-day workshop before the European Safety and Reliability Conference (ESREL) in Dublin, Ireland. Thirty participants from 10 countries attended the workshop. Similar to the previous editions, the attendants were divided into smaller groups for in-depth discussion sessions. In the 2019 and 2021 editions, each group focused on a specific topic. In 2022, in contrast, the three groups focused on the same main issues in an interdisciplinary manner, guided by questions prepared in advance. The goal was to create cross-disciplinary discussions and solutions related to fundamental challenges associated with SRS of autonomous systems.

The preceding sections summarize the discussions according to their main topic. This report does not present each group discussion individually but instead merges the groups' insights about the same issues as an attempt to represent the view of all IWASS participants. While solving issues concerning autonomous systems SRS during a single workshop is not realistic, the findings are a path toward the safe development and operation of autonomous systems for researchers, developers, and regulatory agencies.

---

[1] Proceedings to the 1st International Workshop on Autonomous Systems Safety. Trondheim – Norway, 11-13 March 2019. Available at: https://bit.ly/2SsPrLd
[2] Proceedings to the International Workshop on Autonomous Systems Safety 2021. 20, 21 and 28 March 2021. Available at: https://bit.ly/3jhVvTO

# Findings of IWASS 2019 and 2021

IWASS 2019 and 2021 consisted of discussions around specific themes in a multi-disciplinary approach. The first edition was dedicated to challenges and issues in: *i) Autonomous transportation technology: Society and Individuals in the loop*; ii) *Methods for Safety, Reliability and Security Modelling of Autonomous System*, iii) *System Verification, Processes and Testing,* and iv) *Autonomous Systems' Intelligence and Decision Support*. The second edition focused on similar issues concerning i) *Demonstrating the Safety of Autonomous Systems – verification, validation and risk acceptance*, ii) *Human-on-the-loop – the role of humans in the autonomous system*, iii) *Modeling and simulation for understanding complexity and cascading failure*, and iv) *Artificial Intelligence and Data Analytics in Resilient Autonomous Systems*. The below summarizes the take-home messages from IWASS 2019 and 2021, and the reader is encouraged to access the proceedings for in-depth discussions.

Autonomous systems add to the *complexity* of a system, which also increases the complexity concerning SRS assessment and assurance. Many existing methods are inadequate and lack integrated modelling of hardware, human and software of autonomous systems. Self-learning systems and data quality for training further add complexity to these systems SRS assessment. Therefore, new modeling techniques are required that capture interdependencies and connections. Simulations may assist in this assessment and provide input to decision-making during design and operation.

*Cyber security* and *software risk*s differ from traditional security issues and hardware failures. In these cases, past behavior cannot be used to predict future behavior, particularly in situations where the systems continuously learn.

*Verification and testing* will play a vital role in the development of autonomous systems. While regulatory, ethical and societal requirements need to be addressed, it is of concern how to derive these requirements. Especially, a self-learning (artificial intelligence based-) system needs continuous and integrated verification processes. The results of any verification processes should be communicated openly to the public and regulators to build trust.

Indeed, *Demonstrating the Safety of Autonomous Systems* to show that residual risks are acceptable requires evidence that their performance assessments have been *verified* and *validated*. IWASS explored the topic from three perspectives. First, the machine-centric verification and validation. Technical specialists are by nature optimistic making machine-centric verification and validation more challenging. Second, the human-machine interface verification and validation. A consensus emerged that the current methods for verification and validation in this realm are inadequate. Social requirements verification and validation present an additional challenge. For instance, how the "rules of the road" are established, i.e. parameters measured or mandatory system requirements.

Autonomous systems will be deployed in an open and public environment whose actors and conditions may be changing, sometimes slowly, sometimes rapidly. How can we be sure that societal aspects are adequately reflected during verification and validation of our models?

Concerning *Human-on-the-Loop*, the extent of human involvement in autonomous systems operations and its safety impact is not well established. Levels of Autonomy (LoAs) are likely to oversimplify the role of humans in higher LoAs. While the task load may be reduced in higher LoAs, the tasks may demand a significantly higher level of attention and effort, while being safety critical for the system. The approach to LoAs must thus be revisited for clarifying the human role. Risk Analyses analyze the functions in different operational modes, including any shifts in the LoA and shared control with the human operator.

Autonomous systems' features such as *complexity and possible cascading failures* pose several challenges concerning risk assessment. There is a need for a "framework" with methods for identifying, analyzing and evaluating different hazards, hazardous events and the associated risks. Such a framework should feature qualitative and quantitative methods and should promote the combination and application of both simulation as well as more traditional "discrete logic". Furthermore, the complexity of autonomous systems could be addressed through the compartmentalization (modularization) of a system. However, defining the sub-systems' boundaries and the correct integration of the sub-models with each other remains a challenge.

Autonomous systems are expected to rely on *Artificial intelligence (AI) and data analytics*. AI and data analytics can be applied to autonomous systems in two ways: firstly, it can be applied as part of the systems' intelligence, i.e., information processing, decision-making, or motion control. Secondly, it can be used as part of the verification and safety assurance process of autonomous systems. Whatever the purpose of the application AI is, domain knowledge, interdisciplinary viewpoints and reliable data should be combined. The AI methods used need to suit its purpose in an autonomous system and should be combined with other suitable methods for control of the system. For example, agent-based approaches may be used alongside learning-based systems to incorporate a level of explainability.

# IWASS 2022

# Discussion summary

# *Safe* Autonomous Systems

An increasing level of autonomy can be sought to develop more efficient, cleaner, and economic systems, replace humans in dangerous tasks, or perform tasks that are impossible with non-autonomous technology. Independently of the benefit added by autonomy, these systems must be, above all, *safe*. This section summarizes the general points that were discussed during IWASS on fundamental safety-related questions: *what is safe?* and *how safe is safe enough?*.

First, the sub-section "Autonomous System Operation: Where, When, and How?" discusses the conditions under which autonomous systems are designed to operate and how they can impact safety. The following sub-section, "How Safe Should an Autonomous System Be?" discusses safety levels, risk acceptance, and associated challenges. These points are summarized in the "Considerations for Safe Autonomous Systems", along with additional points on needed elements for a system to be considered *safe*.

## Autonomous systems operation: where, when, and how?

A safe system can be defined as one that does not cause hurt, injury, or loss to human lives[3] above a "tolerable" level. Many industries adopt this general definition as a foundation for a more application-specific one. For instance, a definition of Maritime Safety is "the protection of the crew and passengers aboard vessels, as well as those living or working near bodies of water, from hazards and the risk of injury or fatality". The safety concept may be understood in a broader sense. For instance, a traffic accident involving only autonomous[4] vehicles with no driver or passengers onboard may not directly cause harm or loss of human lives. However, it may indirectly impact lives through a traffic disruption that prevents evacuation from a hazardous area or access to a hospital.

Several elements of an autonomous system relate to safety: reliability of hardware, software, and humans, cybersecurity, connectivity aspects, among others. An essential factor for assessing these elements is the conditions in which the system should operate. An "operational envelope"[5], or "operational design domain" (ODD[6]), can define the geographical limits for the system's operation (e.g., urban areas or rural areas), weather-related aspects (e.g., restricted operation under snow or heavy rain), and others[7]. The operational envelope indicates the conditions under which the system is expected to operate *safely* by design. For instance, if a system's ODD includes nighttime, the perception elements such as cameras, radars, and sensors must detect and correctly classify objects under limited light conditions.

---

[3] Based on Merriam Webster dictionary - https://www.merriam-webster.com/dictionary/safety

[4] The car industry applies the term Automated Driving Systems (ADS) for the software and hardware that perform the driving tasks. For consistency with other industries that apply the "autonomous" term (e.g. autonomous vessels), we will use "autonomous cars" rather than ADS-equipped vehicle.

[5] More information can be seen at the ISO/TS 23860 standard.

[6] More information on ODD for ADS can be seen at SAE J3016 standard

Similarly, if the ODD of a land transport system includes urban areas, the system must correctly apply the rules of the road (for that specific jurisdiction) and be able to deal with the various road elements and road users. Yet, recent events indicate that systems may behave differently than expected even within their ODD. For example, consider the operation of robotaxis (Level 4 Automated Driving Systems – ADS[8]) in San Francisco, allowed under a restricted ODD since 2021. In June 2022, more than a half dozen robotaxis stopped in the middle of a street, blocking traffic for a couple of hours until employees manually moved the autonomous vehicles[9]. Months later, in September, a robotaxi halted on the streetcar tracks, leading the 140 passengers riding the San Francisco N line to be stuck in place for seven minutes before a company employee arrived and moved the vehicle[10]. While these incidents have not led to direct injury or loss of lives, a similar behavior may lead to more severe consequences in other contexts. It is clear from these situations that a simple "stop" operation is not always the safest or most desirable way for the system to fail. As a result, developers are urged to consider what "fail-safety" means in a given situation for a given system. In the above case, it may have been less disruptive if the robotaxis were able to navigate to a parking space before stopping completely.

The robotaxis incidents are a clear example of a system not operating as expected within its defined ODD: vehicles should not stop on the road or over the city streetcar tracks and disrupt traffic. Thus, the operational envelope is implicitly or explicitly accompanied by the system's expected behavior within the prescribed conditions. In addition, systems may breach their ODD – due to unpredicted events or systems' failures, for instance, and their expected behavior during a breach must be assessed. Those systems with higher levels of autonomy are expected to be able to recognize the breach and safely handle the situation. For example, if heavy rain impacting visibility hinders a vehicle perception system and is thus outside of its ODD, a Level 4 vehicle should autonomously recognize the adverse weather and safely achieve a stable stopped condition. Conversely, for systems with a lower level of autonomy, a human operator monitoring and controlling the system - onboard or remotely – may be responsible for handling ODD breaches and hazardous situations within the ODD. A similar approach can be taken for systems designed with high autonomy but for which the operations must include a human-on-the-loop due to legislation requirement. During possibly hazardous situations, thus, the system can hand over control to a human operator.

---

[7] Operational Envelope (OE) and ODD as defined by ISO/TS 23860 and SAE J3016 differ. OE is inspired by ODD definition, but also includes operations under human control.

[8] In the Level 4 of driving automation, the ADS should be able to perform all dynamic driving tasks under specified ODD.

[9] TechCrunch, 2022. https://tcrn.ch/3y6uPcp

[10] SLATE, 2022. https://slate.com/technology/2022/12/san-francisco-waymo-cruise-self-driving-cars-robotaxis.html

Incidents with severe consequences exemplify the complexity of shared responsibilities between autonomous systems and humans. In the first recorded case of a pedestrian fatality involving an ADS-equipped vehicle, in 2018, the car was operating under auto-pilot and detected the pedestrian only 5.6 seconds before the crash. It did not correctly predict her path or reduce the vehicle's speed. The system's design relied on the vehicle operator to take control of the vehicle, but the driver was looking at her phone and failed to take control and avoid the collision[11]. This incident involved an aspect common to other incidents happening over task-switching scenarios; when a system operating autonomously requires human control, often, the time for operators' reaction is not sufficient. Other factors add to the challenge of shared responsibilities. For instance, operators of an autonomous vessel working from an onshore control center may miss sensorial aspects of the ship, reducing their situational awareness; or the possible long monitoring times of a highly automated system may lead to operators' distraction. Additional factors impacting operators' timely (and correct) reaction include automation complacency, overtrust, automation transparency, and others. If systems are designed such that their operation requires support and intervention, humans' capabilities and limitations must be adequately considered. This consideration is particularly important in safety-critical situations of ODD breaches or systems' failure. Achieving a high level of system explainability may also help to ensure that the humans react appropriately in such situations.

## How safe should an autonomous system be?

Naturally, several scenarios involving systems ranging from transportation to the energy field may lead to some type of loss or injury to humans, i.e., the system may behave unsafely in its lifetime (due to inherent or external factors). The discussions about a safe system concern, thus, an *acceptable* level of risk of the system's operation, or *how safe* autonomous systems should be. Some industries have discussed the safety of an autonomous system compared to its human-operated counterpart. Under this perspective, for instance, an autonomous vehicle should be as safe as a human-driven one, i.e., it should not increase the current rate of harm or loss of lives caused by human-driven vehicles. Yet, the "as safe as current systems" perspective raises some issues.

The first issue is that historical data for autonomous systems is insufficient for comparison with human-driven ones. It will take an enormous amount of time for autonomous cars to drive a number of miles sufficient for statistical analysis and accurate comparison with human-driven ones. An approach suggested for bypassing – or complementing – the "miles driven" issue is a behavior comparison: an autonomous system should behave similarly to a human-operated one, e.g. in the automotive domain the UN Regulation No.157[12], which provides in Annex 3a "Component and careful human driver model" that can be used for comparing the behavior of the human driver to the behavior an (autonomous) adaptive lane keeping system. However, characterizing the behavior of a human operator is not always straightforward. It may be simpler to describe the "mean behavior" of highly-trained operators of procedure-based systems such as nuclear power plants. The same cannot be said about other systems, e.g.: car drivers' behavior can highly differ based on age, experience, country, and others. To benchmark autonomous cars' behavior against a human-driven one thus becomes very challenging.

[11] National Transportation Safety Board. 2019. Collision Between Vehicle Controlled by Developmental Automated Driving System and Pedestrian, Tempe, Arizona, March 18, 2018. Highway Accident Report NTSB/HAR-19/03. Washington, DC.

A second issue for the "as safe as current system" approach is that it may lead to ignoring the "unknown unknowns" of autonomous systems' operation. More than replacing a human operator with autonomous perception and decision-making software and hardware, autonomous systems comprise new elements that may lead to safety concerns. These include connectivity issues, emergent failures from unsafe interaction between sub-systems, security concerns related to hacking and spoofing, and others. Consequently, the risk profile of autonomous systems is challenging to compare with current systems. For instance, the number of "daily accidents" (expected loss) might be lower, but the catastrophe potential (maximum loss) higher.

Finally, one of the reasons for developing more autonomous systems is to increase the safety level. Thus, designing systems "as safe as current systems" may limit the benefits that can be achieved with technology: developing systems that are *considerably safer* than current ones. In addition to contributing to a safer society, reducing harms or loss to human lives, systems that are safer than current ones also increase users' trust. Safety must be *communicated* and *demonstrated* to society. *Safer* systems can increase public acceptance of new systems, particularly those for which the public's tolerance concerning accidents involving an autonomous system is lower than for a human-operated one. Equipping autonomous systems with the ability to explain why they have taken specific actions and made certain decisions can also help with this trust element.

## Considerations for a safe autonomous system

The sections above indicates that there is no one solution to the question of how safe an autonomous system should be. Further, the risk tolerance may differ between industries depending on the regulations, stakeholders' acceptance and trust, and current human-operated counterpart risk levels. Nonetheless, some general requirements can be drawn for a system to be safe:

1. The system elements need to be able to be adequately and robustly tested and analyzed.
   a. The system needs to be structured in a modular fashion so that its components can be isolated and "interrogated". Ideally, individual components should be able to provide useful explanations about the decisions that have been made.
   b. The decision-making and control systems should be verifiable, regulated, transparent, and explainable.
   c. The limitations of the technologies used for situational awareness should be identified and understood.

2. An operational envelope and systems' expected behavior must be defined.
   a. A restricted envelope can help to deploy the systems and assess their compliance with the expected behavior before a full deployment as a replacement or complement of current systems.
   b. The possibility of breaching the operational envelope cannot be ignored, and the system's expected behavior in these cases needs to be well described and assessed. Specifically, these systems should have the capability to fail safely.

3. Operations that include human operators in different roles – sharing tasks by design or taking over in hazardous situations - must consider humans' capabilities and limitations.
   a. The design of the operation cannot be driven only by the technological capabilities (e.g., the technology can autonomously handle situations *a* and *b*, and in case of *c* an operator should intervene) or legislation requirements (e.g., legislation requires an operator for liability reasons).
   b. Human operators might be assigned the role of preventing situations that potentially disrupt autonomy (e.g., keeping the autonomous system within its design base envelope).
   c. The design must consider the human capabilities carefully with respect to factors like available response time and ability to perceive and process available information in an event that requires human interaction. The capabilities of humans for non-autonomous systems can be much higher than for autonomous systems as the exposure to some tasks (like responding to something unexpected) may be reduced through less involvement in normal operation.

4. Autonomous systems are often described by their Level of Autonomy, or Degree of Automation. Those generally assume a linear progression of autonomy and may lead to a false perception of simplicity and safety in task-switching between humans and system.

5. Safety is not binary, rather, it has different levels. The more objective answer to "how safe a system is" concerns *risk* levels, which should be assessed against pre-defined acceptable levels.
   a. The "as safe as current systems" approach to autonomous systems safety poses issues for i) objectively comparing systems, and ii) possibly missing the opportunity of developing *safer* systems than current ones.
   b. Safer systems can lead to higher public acceptance of autonomous systems.
6. Safety needs to be adequately and accurately *assessed*, *demonstrated,* and *communicated*.
7. Risk analysis is a good tool for assessing a system's safety. Risk specialists/risk analysts must be engaged early in the design process together with specialists from other disciplines.

---

[12] UN Regulation No 157 – Uniform provisions concerning the approval of vehicles with regards to Automated Lane Keeping Systems [2021/389] (OJ L 82 09.03.2021, p. 75, ELI: http://data.europa.eu/eli/reg/2021/389/oj)

# Risk Assessment Methods and Safe Autonomous Systems

Risk-based design is one approach to building safe systems from the early design phases. Qualitative, quantitative, and simulation methods are used to assess the system's performance in different operational scenarios. The design is consequently modified to remove identified system weaknesses. However, it is to be noted that there is a difference between assessing a risk level through risk assessments and safety assurance. The latter concerns the demonstration that the system behavior and performance correspond with the risk assessment assumptions. Risk assessment and safety assurance may be carried out by different entities.

The foundations of risk assessment for autonomous systems will generally not differ much from conventional (non-autonomous) systems. Hence, currently used methods represent a good starting point for analyses and can benefit from new guidance words for better adaptation to autonomous systems. Yet, method's limitations must be recognized and addressed before their use for decision-making. Depending on the method, these limitations may concern the complexity of autonomous systems and the identification of complex hazards arising from interaction or AI-based systems that may lead to unanticipated behavior. However, new methods may also give new insights and should be applied together with existing ones.

Regarding quantitative risk assessment, the current approach to identifying hazards and developing scenarios, assessing their frequency, and the associated consequences should still be applicable. Related challenges include identifying failures and interactions that may lead to undesired consequences, and the data for quantifying frequencies and consequences. Simply relying on historical accidents and experience of these systems is insufficient. Yet, data from similar systems are sometimes available and often not used. For example, there is much experience with remote monitoring systems through control room environments from the nuclear industry. This data can be leveraged for risk assessment of autonomous systems designed for remote control and monitoring while recognizing the differences between the systems' complexities.

Algorithms and AI-based systems can fail in many different ways, sometimes due to small scenario changes. Processing the many risk scenarios related to autonomous systems operations is a resource-consuming task that may become almost impossible. This challenge raises the question of whether scenario-based approaches are still applicable and relevant for assessing autonomous systems.

Additional aspects concerning risk assessment for autonomous systems relate to the many sub-systems involved and the range of competencies and disciplines that should be applied. Consider, for instance, human error: while many autonomous systems aim at replacing the human operator, they will still be part of the operation in the near to medium future. The role of humans in autonomous systems may be monitoring, indirect or direct control, troubleshooting, or merely passenger support. Human error can, thus, still impact the system's safe operation. Unfortunately, current methods for assessing human error through Human Reliability Assessment (HRA) or improving human performance through Human Factors Engineering often do not account sufficiently for automation and the problems that may arise from its use. Those methods may be better suited for simple, repetitive tasks that humans interacting with autonomous systems will no longer perform. Nevertheless, many aspects of automation and autonomy's impact on human performance have been widely discussed, such as skill degradation and automation complacency. More recently developed cognitive-based HRA methods can also provide a suitable foundation for assessing human error in autonomous systems operations.

# New approaches to risk assessment of autonomous systems

An often-discussed topic in risk assessment for autonomous systems is the possible need for new and adapted methods. Academia has produced methods and approaches that can handle a high degree of complexity. However, clear criteria for when an existing method is adequate for an analysis and when a new one is needed are still to be developed. These criteria must indicate to the industry the benefits of any new methods, as their application will initially require resources in terms of training, knowledge, and changes in the safety assessment processes. Authorities, regulators, or third parties must also accept the new methods as needed and sufficient.

Some recently proposed methods are bio-inspired[13], which allows an autonomous system to understand when it fails. This can be used in a simulated environment to assess scenarios. However, such an approach may not work for all types of autonomous systems and is dependent, among other factors, on the Level of Autonomy. Another approach inspired by mathematics is agent-based models with a game theoretical approach[14].

AI-based methods, where the AI learns from data to facilitate the assessment, are also seen as promising tool[15]. AI could help identify scenarios that may be overlooked by combining system knowledge and knowledge of earlier assessments. AI-based methods need to be informed, meaning that the models should be built with domain knowledge and the data is used to tune the model. In some AI techniques, for example deep learning, little is known about the parameters. Hence, the models should be transparent and trustworthy, i.e., it should be possible to assess if something is deviating from the expected and if the results are reliable.

# Considerations on risk assessment for autonomous systems

The Risk-related disciplines – risk assessment and risk management, reliability, security – provide several methods for achieving safer autonomous systems. While the complexity added by those systems challenges current approaches, the foundations and the accumulated knowledge and experience acquired through the development of Risk Sciences allow to draw general requirements and insights:

---

[13] Ventikos NP, Louzis K. Developing next generation marine risk analysis for ships: Bio-inspiration for building immunity. Proceedings of the Institution of Mechanical Engineers, Part O: Journal of Risk and Reliability. 2022;0(0). doi:10.1177/1748006X221087501

[14] Ramos, M.; Moura, M.; Lins, I.; Ramos, F. The use of Game Theory for Autonomous Systems Safety: An Overview. Proceedings to ESREL 2021.

[15] Hegde, J.; Rokseth, B. Applications of machine learning methods for engineering risk assessment – A review, Safety Science, 2020. https://doi.org/10.1016/j.ssci.2019.09.015.

1. Validation and Verification (V&V) should be performed earlier in (and throughout the whole) the system design process to provide feedback.
   a. V&V must cover software aspects and testing.
   b. Additional competence requirements may be needed for people involved in the system development process.
   c. Consideration must also be given to demonstrate that the system complies with relevant standards and regulations. This includes the development processes and V&V approaches used.
2. Environmental conditions may strongly impact the safety of several systems. Risk assessment cannot rely solely on historical data for those conditions, given climate change around the globe.
3. Safety needs to be continuously evaluated throughout a system's lifetime.
   a. Evaluations should consider lessons learned and changes of operational, environmental, and organizational conditions.
   b. The large quantity of data generated by connected systems must be leveraged for updating risk assessments throughout the time.
   c. Monitoring and verification at runtime can be used to continuously evaluate the system and its relevant subsystems.
4. Cybersecurity may be complex for autonomous systems compared to conventional systems, yet it has to be explicitly incorporated into the risk assessments.
   a. The reliance on communication technologies to inform humans about the status of the autonomous system may lead to abusable vulnerabilities.
   b. For example, shipping is vital for global trade and supply chains, autonomous ships could become the target of (state) terrorism to disrupt the supply of individual countries
   c. Autonomous systems are more reliant on sensors and raw environmental data than classical ones since this data is what informs the decisions that are made. As such multiple sensors may be incorporated that measure the same thing so that the system can identify if one or more of these sensors has been maliciously compromised.
5. People are still involved in the design of autonomous systems. Therefore, they need to understand the risks that are associated with the system. Conducting useful and high-quality risk assessments of autonomous systems, thus, requires the right team with a shared understanding and language of work.
   a. In the context of autonomous systems, several disciplines need to be involved in the assessment.
   b. Experts from computer science alone are not enough to assess the impact of an autonomous system. Experts in other disciplines that

focus for example on the use of the system are required to be involved in the assessment. Domain experts are necessary.

c. Regulators could define which disciplines should be involved in the assessment.

d. An approach that identifies and includes the most relevant stakeholders and disciplines in the risk assessment must be developed and applied.

# Regulating Autonomous Systems: How to Consider Safety?

Every industry and every country has its own regulations regarding the requirements for autonomous systems development, testing, and deployment. Standardizing these requirements concerning what constitutes a safe autonomous system can provide essential guidance for regulators or appointed assessors. Standards that are adopted by various countries can particularly benefit autonomous systems that cross these geographical borders, such as international cargo transport by autonomous ships. For berthing operations, the same standards and regulations need to apply to make these operations feasible.

On an international level, adopting international regulations is challenging, as each industry has different approaches to international regulation. For example, the International Maritime Organization (IMO) only provides recommendations for adoption to member states. Thus, an autonomous vessel developed according to the international recommendation may not be accepted in a country that did not adopt this resolution.

In addition to standards on systems' development, regulators need to set requirements regarding risk assessment or safety cases as part of the assurance process. A balance needs to be found between extensive and coarse analysis and requirements for an autonomous system. Regulations may in this way impact education, university programs and industry perception of the importance (or lack of importance) of interdisciplinarity and Reliability, Availability, Maintainability and Safety (RAMS) competencies.

Some examples demonstrate the interest of regulators in autonomy projects. For instance, the UK railway regulation gets involved in an early project stage to understand the goals and requirements of an autonomous system. Similarly, the Norwegian Maritime Authority has joined several projects on autonomous cargo and passenger vessels to understand the challenges regarding the current regulations and how a satisfactory solution may be reached. However, in other countries or industries, the regulators cannot be involved so that they can remain impartial and not favor specific solutions, which may create gaps in systems and technology knowledge.

Some of the challenges the regulators currently face include:

- Risks and responsibilities associated with AI and decision systems;
- Human role in autonomous systems and acceptable automation;
- The fast pace of development and adoption of new technologies, which is outpacing the regulatory processes
- Acceptance and risk assurance criteria as to what constitutes a safe system and if fully autonomous systems should be better than human controlled systems.

Ethical aspects complement the above challenges: when it comes to decision-making, autonomous systems may not meet requirements regarding ethical expectations. Determining such requirements is not straightforward: ethics is not an objective and quantifiable topic. Utilitarian ethics may be preferred insofar they more easily translate into cost-benefit analyses, but this would leave outside other ethical theories that might be equally, if not more, relevant. Hence, some level of ethical requirements and expectations must need to be defined by the regulators. For instance, it is recognized that several equity issues are part of our society. As a reflection of society, biases may find their way into the decision algorithms. Databases used for training (AI-based) decision systems need to be checked for such biases, and regulators must decide if they are acceptable to a certain level. Programmers will be responsible for the algorithms and together with system designers are selecting training data and desired behavior. Thus, ethical should be also addressed by them. Different area of the world may also have different approaches towards ethical requirements. The European Commission issued for instance Ethics Guidelines for Trustworthy Artificial Intelligence (AI) in 2019[16] that structured as key requirements the strategy to pursue a -human centric AI. In summary, ethical development and deployment of AI is required, meaning that guidelines for these processes need to be provided to the industries.

Ethical issues are often closely discussed with legal matters, such as the implications of AI failures, misuse, or liability concerns. These close ties between the topics have sometimes been seen as a barrier for deploying highly autonomous systems. A recorder or so-called "black box" for autonomous systems could be required to clarify legal issues, such that an accident can be fully explained, and errors, faults, and liabilities can be thoroughly investigated.

---

[16] HLEG, AI. "Ethics guidelines for trustworthy AI. European Commission High-Level Expert Group on AI, April 8." (2019).

# IWASS 2022: Final message

The trend towards higher autonomy in a diverse range of systems and operations is crucial for enabling new types of land-based and maritime transportation, enhanced mapping and monitoring of oceans and areas on land, and advanced inspections of physical structures difficult to access. Autonomous systems may be a step towards safer and more efficient operations. Still, more software and advanced control systems lead to complexities and risks that are challenging to identify, analyze, evaluate, monitor, and mitigate. The third IWASS workshop confirms that the "key" to solving the SRS challenges has yet to be found. Nevertheless, we identified the critical areas where more effort is needed:

- What is a safe system? Even with highly autonomous systems, their designs must be user-centered, enabling humans to intervene timely and safely. Furthermore, risk specialists must be engaged early in the design process, which is not always the case.
- What is safe enough for an autonomous system? Risk acceptance remains a challenge in which there are several proposals for solutions, including different types of safety envelopes and constraints.
- The methodology for analyzing and evaluating risks of software-intensive systems is advancing. Still, main challenges remain, for example, with respect to balancing comprehensiveness and efficiency, and for performing risk assessments for systems in operation. Enabling technologies in terms of improved and cheaper sensors and computers enhance the possibility for simulation, use of digital twins for more precise predictions of system and operational performance.
- Levels of autonomy introduce different risks. In risk analysis of autonomous systems, it is important to analyze the functions in different operational modes, including any shifts in the level of autonomy and shared control with the human operator.
- Validation and verification (V&V) of autonomous systems also remains a challenge, which is closely linked to the methodological problems of assessing the risk. If the methodology for risk assessment is improved, V&V becomes easier, for example, in selecting and prioritizing risk scenarios in an assurance process. V&V efforts need to be trustworthy and acceptable, requiring high quality.

Finally, a gap needs to be reduced concerning disseminating theory and recommendations from academia on which approaches should be recommended when and where, and sharing knowledge, needs, and experience from industry and regulators the other way. In this sense, as part of IWASS effort, there will be input on the ISO/TC 159/SC 4 to disseminate this knowledge through updated standards.

Ergonomics of human-system interaction in a new upcoming standard to address human factors for intelligent and autonomous systems. This effort of reaching out from academia to industry and regulators will be a continuing goal of the past and future IWASS workshops.

# Organizing Committee

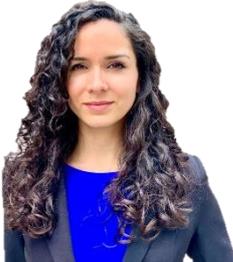

### Marilia Ramos, PhD - UCLA

Dr. Marilia Ramos is a Research Scientist at the B. John Garrick Institute for the Risk Sciences, UCLA, where she manages research projects on risk and reliability. She holds a PhD in Chemical Engineering from the Federal University of Pernambuco, Brazil. She currently develops research on advanced Human Reliability Analysis methods and applications; and leads and collaborates with research projects on human behavior modeling and on risk management for nonindustrial systems, such as autonomous systems operations and wildfire evacuations. Additionally, she is an Instructor of the Human Reliability Analysis course for the Master of Science in Engineering Online Program (MSOL) at UCLA.

www.mariliaramos.net  marilia.ramos@ucla.edu

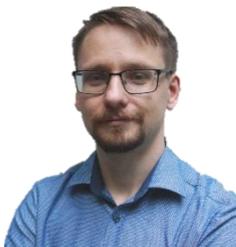

### Christoph A. Thieme, PhD – SINTEF

Dr. Christoph Thieme is a researcher at SINTEF Digital in Trondheim, Norway, where he applies his knowledge to different research projects related to safety and security of socio-technical systems. He obtained his PhD in Marine Technology from NTNU, with specialization in safety, reliability, and risk assessment for autonomous systems. He has experience with risk assessment of autonomous systems with a focus on software safety and human-machine interaction. Additionally, he is a visiting professor at the University of Toulon lecturing on Risk and Reliability engineering and potential application of AI methods, building on his research and the insights gained at the IWASS workshops within safety, reliability, and security for autonomous systems.

www.christophthieme.com christoph.thieme@sintef.no

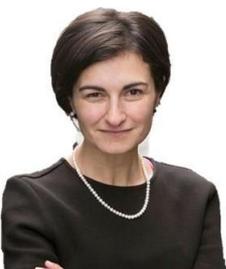

### Maria Chiara Leva, PhD - Technological University Dublin

Dr Maria Chiara Leva (PI in ESHI) is the co-chair of the technical committee on Human Factors for the European Safety and Reliability Association (ESRA), former chair of the Irish Ergonomics Society and Founder of the Human factors in Safety and Sustainability research group in TU Dublin. Dr Leva has more than 60 publications on Human Factors (HF), Operational Risk Assessment and Safety Management in Science and Engineering Journals. She is a Lecturer in TU Dublin and visiting lecturer for Risk Assessment and Safety Management in the School of Engineering, associated PI in the Science and Technology in Advanced Manufacturing research center and in the Centre for Innovative Human Systems in Trinity College Dublin.

https://ie.linkedin.com/in/chiara-leva-2427532 mariachiara.leva@tudublin.ie

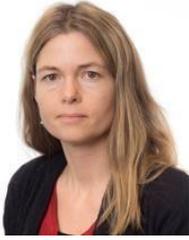

**Ingrid B. Utne, PhD** – NTNU

Dr. Ingrid Bouwer Utne is a Professor at Department of Marine Technology, NTNU, where she performs research on risk assessment and modeling of marine and maritime systems. Utne is an affiliated Researcher in the Center of Excellence on Autonomous Marine Operations and Systems (NTNU AMOS). She is a principal investigator of the research projects UNLOCK and ORCAS. These projects focus on supervisory risk control and bridge the scientific disciplines of risk management and engineering cybernetics aiming to enhance safety and intelligence in autonomous systems.

ntnu.edu/employees/ingrid.b.utne ingrid.b.utne@ntnu.no

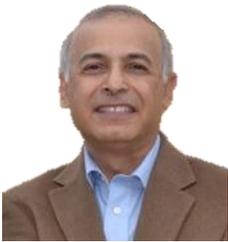

**Ali Mosleh, PhD** – UCLA

Dr. Ali Mosleh is Distinguished University Professor and holder of the Knight Endowed Chair in Engineering at UCLA, where he is also the director of the Institute for the Risk Sciences. He conducts research on methods for probabilistic risk analysis and reliability of complex systems and has made many contributions in diverse fields of theory and application. He was elected to the US National Academy of Engineering in 2010 and is a Fellow of the Society for Risk Analysis, and the American Nuclear Society. Prof. Mosleh is the recipient of many scientific achievement awards.

risksciences.ucla.edu/institute-director mosleh@ucla.edu

# Organizers and Sponsors

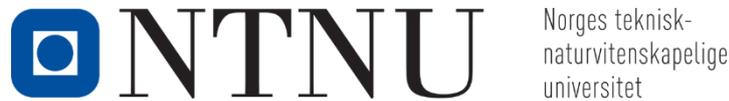

**Department of Marine Technology, Norwegian University of Science and Technology (NTNU), Trondheim, Norway**

The Department of Marine Technology at NTNU provides world-class education and research for engineering systems in the marine environment. The focus is on methods and techniques for sustainable development and operation of ship technology, fisheries and aquaculture technology, oil and gas extraction at sea, offshore renewable energy, and marine robotics for mapping and monitoring the ocean. The Department hosts an excellent research group working on safety and risk management of marine and maritime systems. The Centre of Excellence Autonomous Marine Operations and Systems (NTNU AMOS) is also located at the Department. The Norwegian University of Science and Technology in Trondheim (NTNU) is the largest university in Norway.

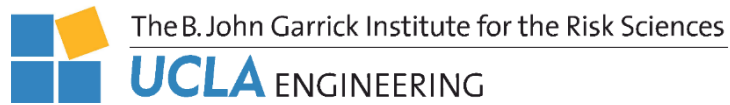

**The B. John Garrick Institute for the Risk Sciences, University of California, Los Angeles, USA**

The B. John Garrick Institute for the Risk Sciences has declared its mission to be the advancement and application of the risk sciences to save lives, protect the environment and improve system performance. The purpose of the Garrick Institute is for the research, development, and application of technology for (1) quantifying the risk of the most serious threats to society to better enable their prevention, reduce their likelihood of occurrence or limit their consequences and (2) improving system performance with respect to reliability and safety. The institute is hosted at the Department of Engineering at the University of California Los Angeles (UCLA).

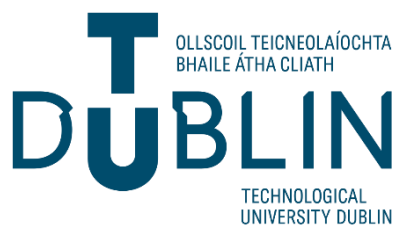

**Human Factors in Safety and Sustainability, Technological University Dublin**

Technological University Dublin, Human Factors in Safety and Sustainability (HFISS) research group.[17] TU Dublin is a leader in STEM disciplines, the University supports the largest cohort of students of business, media, culinary arts, and the creative and performing arts. In TU Dublin the Research group for Human Factors in Safety & Sustainability (HFISS) is an innovative hub of multidisciplinary expertise committed to human centred design and improvement for the safety and sustainability of intelligent complex systems. This research group promotes the consideration of human and organisational factors in sectors where breakdowns between the automated system and the human operator can have fatal consequences. At HFISS, we provide the interdisciplinary skillset for the development of Collaborative Intelligence systems blending expertise in AI with expertise in Human Factors, Human Reliability Analysis, Neuroergonomics and System Safety Engineering.

**DNV**

DNV is a global quality assurance and risk management company. DNV provides classification, technical assurance, software and independent expert advisory services to several industries. Combining technical, digital and operational expertise, risk 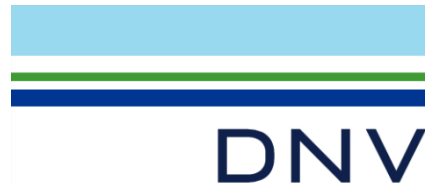 methodology and in-depth industry knowledge, DNV GL assists its customers in decisions and actions with trust and confidence. With origins stretching back to 1864 and operations in more than 100 countries. DNV are dedicated to helping customers make the world safer, smarter and greener.

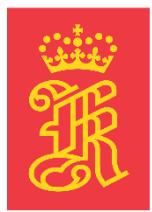

**Kongsberg Maritime**

Kongsberg Maritime (KM) is a leading supplier of offshore and marine energy solutions, deck machinery and automation systems. In addition, KM provides services related to complex system integration, and vessel design. KM is a leader in marine ship intelligence, automation and autonomy and is a part of the Kongsberg Group.

---

[17] https://www.tudublin.ie/research/discover-our-research/research-institutes-centres-and-groups/hfiss/about/

### Research Council of Norway

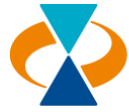

The Research Council of Norway serves as the chief advisory body for the government authorities on research policy issues. The Research Council of Norway co-financed the IWASS workshop through the MAROFF knowledge-building project for industry ORCAS (Project number 280655) and the FRINATEK project UNLOCK (Project number 274441).

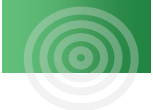

# IWASS 2022 Participants

### Organizing Committee

| Name | Affiliation |
|---|---|
| Ingrid B. Utne | Norwegian University of Science and Technology (NTNU) - Norway |
| Ali Mosleh | University of California Los Angeles (UCLA) - U.S.A. |
| Christoph Thieme | SINTEF Digital - Norway |
| Marilia Ramos | UCLA |
| Maria Chiara Leva | Technical University of Dublin (TU Dublin) - Ireland |

### Session Hosts and Panelists

| Name | Affiliation |
|---|---|
| Claire Blackett | Institute for Energy Technology (IFE) - Norway |
| Mario Brito | University of Southampton - U.K. |
| Jakub Montewka | Gdynia University - Poland |

### Participants

| Name | Affiliation |
|---|---|
| Ammar Abbas | Software Competence Center Hagenberg |
| Andreas Bye | IFE - Norway |
| Andrey Morozov | University of Stuttgart - Germany |
| Caroline Kristensen | SINTEF Digital - Norway |
| Caroline Metcalfe | proactima - Norway |
| Chris Harrison | Rail Safety and Standards Board (RSSB) - U.K. |
| Henrik Bjelland | University of Stavanger - Norway |
| Hector Diego Estrada Lugo | TU Dublin - Ireland |
| Hyungju Kim | University of South-Eastern - Norway |
| John Andrews | University of Nottingham - U.K. |
| Kevin Heaslip | University of Tennessee - U.S.A. |
| Marie Farrell | The University of Manchester - U.K. |
| Mary Ann Lundteigen | NTNU - Norway |
| Niav Hughes Green | U.S. Nuclear Regulatory Commission (NRC) - U.S.A. |
| Nikolaos P. Ventikos | National Technical University of Athens, School of Naval Architecture & Marine Engineering - Greece |
| Ørnulf Jan Rødseth | SINTEF Ocean - Norway |
| Osiris Valdez Banda | Aalto University - Finland |
| Salvatore Massaiu | IFE - Norway |
| Silvia Tolo | University of Nottingham - U.K. |

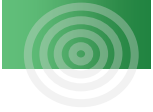

| Name | Affiliation |
|------|-------------|
| Sizarta Sarshar | IFE - Norway |
| Sheng Ding | University of Stuttgart - Germany |
| Tarannom Parhizkar | UCLA - U.S.A. |
| Timo Frederik Horeis | IQZ GmbH - Germany |
| Tunc Aldemir | Ohio State University - U.S.A. |

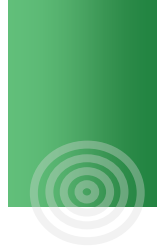